\documentclass[12pt]{article}
\usepackage{epsfig}

\newcommand{\be}{\begin{eqnarray}}
\newcommand{\ee}{\end{eqnarray}}

\newcommand{\ba}{\begin{array}}
\newcommand{\ea}{\end{array}}
\newcommand{\mpp}{m_{\pi\pi}}

\begin{document}
\rightline{RUB-TP2-16/99}

\begin{center}
{\Large {\bf
Hard exclusive electroproduction of two pions}}

\vspace{0.7cm}

B.~Lehmann-Dronke$^{a}$, P.V.~Pobylitsa$^{b,c}$,
M.V.~Polyakov$^{b,c}$, A.~Sch\"afer$^{a}$, and
K.~Goeke$^c$

\vspace{0.5cm}

{\em$^a$ Institut f\"ur Theoretische Physik, Regensburg University,
93040 Regensburg, Germany}

{\em $^b$Petersburg Nuclear Physics Institute,
  188350, Gatchina, Russia}

{\em $^c$ Institut f\"ur Theoretische Physik II, Ruhr-Universit\"at Bochum,
D-44780 Bochum, Germany}

\end{center}

\vspace{0.8cm}

\centerline{\bf Abstract}
\begin{center}
\begin{minipage}{15cm}
We have calculated the leading order amplitude of hard exclusive
production of two pions in lepton nucleon scattering.
At leading twist a pion pair can be produced only
in an isospin one or zero state. We have shown that isoscalar states
are produced predominantly for $x_{Bj}>0.3$ and with an invariant mass
of the two pions close to the threshold (S-wave) and
in the $f_2$ resonance region (D-wave). These
isoscalar pion pairs are mostly produced by two collinear
gluons. Hence, comparing the production of charged and neutral
pion pairs as a function of $x_{Bj}$ and $\mpp$, one can get
information about the gluonic component of two-pion
distribution amplitudes.

\end{minipage}
\end{center}

\vspace{0.3cm}
\noindent {\bf Introduction}
\vspace{0.2cm}\\
Hard exclusive reactions open a new way to study the partonic structure
of hadrons. The factorization theorem \cite{CFS} states that
the amplitude of the reaction:
\be
\gamma^\ast_L + T \to F + T'
\label{processgeneral}
\ee
at large collision energy $W\to\infty$, large virtuality of the
photon $Q^2\to\infty$, fixed $Q^2/W^2$ (Bjorken limit), and with the masses
$M_T,M_{T'},M_F\ll Q$ can be written in the form:
\begin{eqnarray}
   &&
   \sum _{i,j} \int _{0}^{1}dz  \int dx_1\,
   f_{i/T}^{T'}(x_1 ,x_1 -x;t,\mu ) \,
   H_{ij}(Q^2 x_1/x,Q^{2},z,\mu )
   \, \Phi^F_{j}(z,\mu )
\nonumber\\
&&
   + \mbox{power-suppressed corrections} ,
\label{factorization}
\end{eqnarray}
where $f_{i/T}^{T'}$ is a $T\to T'$ skewed parton distribution
\cite{CFS,Dmuller,Ji,Rad} (for a review see \cite{JiReview}),
$\Phi^F_{j}(z,\mu )$ is the distribution amplitude of the hadronic
state $F$ (not necessarily a one particle state), and $H_{ij}$ is a hard
part computable in pQCD as a series in $\alpha_s(Q^2)$.
For the experimentally accessible range of $Q^2$ and $W$ the significance
of this factorization might, however, be
limited by the size of
higher twist effects \cite{Bel2,VMS}.
Let us also note that often NLO corrections in hard exclusive
reactions are noticeable \cite{Bel,MPS,KMP99}. Therefore, in principle,
reliable NLO calculations of the hard part $H_{ij}$ are desirable.

In the present paper we shall study the hard exclusive electroproduction
of two pions to leading order in the strong coupling constant $\alpha_s$.
The related process of photoproduction was studied in
Ref.~\cite{LDMS}. At leading twist the pion pair
can be produced in a state with isospin one or zero.
As argued in \cite{MVP98} at small $x_{Bj}$, i.e. $x_{Bj}<0.1$, the
pions are produced mostly in the isovector state because the dominant
reaction mechanism is
exchange of two gluons with positive C-parity.
Our prime
interest will be
the production of pions in the valence region of Bjorken $x$,
i.e.~$x_{Bj}>0.3$.
In this region
(relevant for CEBAF, HERMES, COMPASS) the production of two
pions is dominated by $q \bar q$ exchange, which leads to a sizeable
admixture of pion pairs with isospin zero
\cite{DGP99}. We shall compute in the present paper
the ratio of the cross sections for $I=0$ and $I=1$ pion pairs
as a function of Bjorken $x$ and the invariant mass $\mpp$ of the
produced pions.

\begin{figure}[t]
\vspace*{1cm}
\centering
\epsfig{file=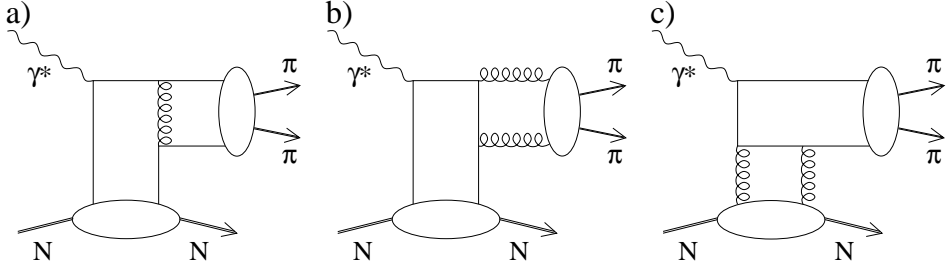, height=3.5cm}
\caption{{\it Typical
leading order diagrams for hard exclusive pion pair
production.}}
\label{graph}
\end{figure}

The interesting feature of $I=0$ production of two pions is that
these pions can originate not only from a collinear $q \bar q$
pair but also from a pair of collinear gluons,
see Fig.~\ref{graph}b). Therefore measurements of
$I=0$ hard pion production could give important information about
the gluon content of the two-pion distribution amplitude.
We predict the most favorable kinematic range
for such measurements.

\vspace{0.3cm}
\noindent {\bf Amplitude of the hard two-pion production}
\vspace{0.2cm}\\
In this section we compute the leading amplitude in $1/Q^2$
of hard two-pion electroproduction.
Due to the factorization theorem for exclusive hard reactions
\cite{CFS} this amplitude can be written as a convolution of
the skewed parton distributions in the nucleon (SPD), the distribution
amplitudes of the produced pions ($2\pi$DA) , and a hard part computable in
perturbative QCD as a power series in $\alpha_s$. Hereby we consider the
processes: \be
\gamma_L^*(q) + B_1(p)\to 2\pi(q')+B_2(p+\Delta)\, ,
\ee
where $\Delta=p'-p$ is a four-momentum transfer to a target
baryon $B_1$, $-q^2=Q^2$ goes to infinity, and $x_{Bj}=Q^2/2 p\cdot q$ is
fixed. The state $B_2$ can be a nucleon or a heavier baryonic state.
Let us note that the corresponding amplitude for a transversally
polarized virtual photon is $1/Q$ power suppressed. Therefore the
considered (sub-) process with the longitudinally polarized photon, for
which the factorization theorem holds, is sufficient to calculate the
leading contribution of the whole electroproduction process
$e^-(k)+B_1(p)\to 2\pi(q')+e^-(k-q)+B_2(p+\Delta)$.

The leading order amplitude corresponding to diagrams of the type
as shown in Fig.~\ref{graph} has the form\footnote{We write the
most general result including the contribution
of graphs of the type Fig.~\ref{graph}c) which are proportional
to the skewed gluon distribution in the nucleon.
It contributes only to the production of pions
with isospin one and is very small in the valence region.
Hence we shall neglect this contribution in the numerical estimates.
For the skewed gluon distribution we use
$F_{G}(\tau,\xi)$ in the
notation of Ref.~\cite{JiReview}.}:
\be
&&\langle B_2(p'), \pi^a\pi^b(q')| J^{\rm e.m.}\cdot
\varepsilon_L|B_1(p)\rangle=
-(e 4 \pi \alpha_s)\frac{N_c^2-1}{N_c^2} \frac{1}{4 Q}
\int_{-1}^1 d\tau \int_0^1 dz\nonumber\\
\label{P-general}
&\times&\Biggl[\sum_{f,f'}
F_{ff'}(\tau,\xi) \Phi_{f'f}^{ab}(z,\zeta,m_{\pi\pi})
\Biggl\{ \frac{e_{f'}}{z\bigl(\tau+\xi \bigr)-i0}+
\frac{e_{f}}{(1-z)\bigl(\tau-\xi \bigr)+i0}
\Biggr\}\nonumber\\
&-&\frac{2 N_c}{N_c^2-1}\sum_f e_f
F_{ff}(\tau,\xi) \Phi_{G}^{ab}(z,\zeta,m_{\pi\pi})
\frac{1}{z(1-z)}
\Biggl\{ \frac{1}{\bigl(\tau+\xi \bigr)-i0}-
\frac{1}{\bigl(\tau-\xi \bigr)+i0}
\Biggr\}\nonumber\\
&+&\frac{4 N_c}{N_c^2-1}\sum_f e_f
\tau F_{G}(\tau,\xi) \Phi_{ff}^{ab}(z,\zeta,m_{\pi\pi})\frac{1}{z(1-z)}
 \frac{1}{\bigl[\tau+\xi -i0\bigr]
\bigl[\tau-\xi +i0\bigr]}
\Biggr]
\, .
\ee
Here $e_f$ is the charge of a quark of flavor $f=u,d,s$ in units of
the proton charge ($e_u=2/3,\; e_d=e_s=-1/3$),
$\xi$ is the skewedness parameter (see definition below), and $\zeta$
is the longitudinal momentum fraction carried by the pion with flavor $a$.
The longitudinal photon polarization vector is chosen as:
\be
\varepsilon_L=
\frac{1}{Q} (q^3,0,0,q^0)
\, .
\label{polarization}
\ee
The function $F_{ff'}(\tau,\xi)$ is a skewed quark distribution
defined as:
\be
F_{ff'}(\tau,\xi)=
\int \frac{d\lambda }{2\pi }e^{i\lambda \tau}
\langle B_2(p^{\prime })|T \bigl\{
\bar \psi_{f'}
(-\lambda n/2){\hat n}
 \psi_f (\lambda n/2)\bigr\}
|B_1(p)\rangle  \, ,
\label{SPD1}
\ee
with $\hat n=\gamma_\alpha n^\alpha$ and the light cone vector $n$
normalized so that:
\be
n^2 &=& 0,
 \hspace{2cm} n\cdot (p + p') \;\; = \;\; 2 \, .
\label{n-normalization}
\ee
The quark two-pion distribution amplitude $\Phi^{ab}_{f'f}(z,\zeta)$
is defined as:
\be
\Phi_{f'f}^{ab}(z,\zeta)=
\int \frac{d\lambda}{ 2\pi} e^{-i\lambda z (q'\cdot n^\ast)}
\langle \pi^a \pi^b |T \bigl\{
\bar \psi_{f}
(\lambda n^\ast){\hat{n}^\ast}
 \psi_{f'} (0)\bigr\}
|0\rangle
\label{DAq}
\ee
and the two-pion gluon distribution amplitude $\Phi^{ab}_G(z,\zeta)$
as: \be
\Phi^{ab}_G (z, \zeta, m_{\pi\pi}^2 ) =
\frac{1}{n^\ast\cdot q'}
\int \frac{d\lambda}{ 2\pi} e^{-i\lambda z (q'\cdot n^\ast)}
\langle \pi^a \pi^b | \bigl\{
n^{\ast\mu} n^{\ast\nu} \,
G^A_{\alpha\mu}(\lambda n^\ast) G^{A\alpha}_{\nu}(0)
\bigr\}
|0\rangle  \, ,
\label{DAg}
\ee
where for convenience we introduced another light cone vector
$n^\ast$ such that
$n\cdot n^\ast=1$
and $n^\ast\cdot n^\ast= 0$. The variable $z$ is the fraction
of the longitudinal (along the vector $n$) meson pair momentum
$q'=q-\Delta$ carried by one
of the quarks or gluons respectively. The variable $\zeta$ characterizes the
distribution of the longitudinal component of $q'$ between the two pions:
\be
\zeta= \frac{(k^a\cdot n^\ast)}{(q'\cdot n^\ast)}\, ,
\ee
where  $k^a$ is the momentum of the pion $\pi^a$.

The vectors $n$ and $n^\ast$ are
linear combinations of the photon, initial,
and final baryon momenta:
\be
\nonumber
p^\mu&=&(1+\xi)n^{\ast\mu}+
(1-\xi )\frac{\bar M^2}{2} n^\mu -\frac 12 \Delta_\perp^\mu\\
\nonumber
p^{\prime \mu}&=&(1- \xi)n^{\ast\mu}+
(1+ \xi)\frac{\bar M^2}{2} n^\mu +\frac 12 \Delta_\perp^\mu\\
\nonumber
q^{\mu}&=&-2 \xi n^{\ast\mu}+
\frac{ Q^2}{4\xi} n^\mu\\
\bar M^2&=&\frac 12
\bigl(M_{B_1}^2+M_{B_2}^2-\frac{\Delta^2}{2} \bigr)\, .
\label{kin}
\ee
The transverse plane is a plane orthogonal to
the plane defined by the light-cone vectors $n$ and $n^\ast$.
The skewedness parameter $\xi$
(the longitudinal component of the momentum transfer) is defined as:
\begin{equation} \xi =-\frac 1 2 (n\cdot \Delta ).
\label{xi-def} \end{equation}
In the Bjorken limit $Q^2\to \infty$, the
skewedness parameter $\xi$ is expressed in terms of Bjorken $x$ as
$\xi=\frac{x}{2-x}$.

The two-pion distribution amplitudes (\ref{DAq},\ref{DAg})
can be flavor decomposed as: \\
\underline{$\pi^+\pi^-$ production}:
\be
\Phi^{f'f}_{\pi^+\pi^-}(z,\zeta, m_{\pi\pi})&=&
\delta^{f'f} \Phi^{I=0}(z,\zeta, m_{\pi\pi})+
\tau_3^{f'f} \Phi^{I=1}(z,\zeta, m_{\pi\pi})\, ,\\
\Phi^{\pi^+\pi^-}_G(z,\zeta, m_{\pi\pi})&=&\Phi^G(z,\zeta, m_{\pi\pi})\, ,
\label{DA2}
\ee
\underline{$\pi^0\pi^0$ production}:
\be
\Phi^{f'f}_{\pi^0\pi^0}(z,\zeta, m_{\pi\pi})&=&
\delta^{f'f} \Phi^{I=0}(z,\zeta, m_{\pi\pi})\, ,\\
\Phi^{\pi^0\pi^0}_G(z,\zeta, m_{\pi\pi})&=&\Phi^G(z,\zeta, m_{\pi\pi})\, ,
\label{DA3}
\ee
where $\Phi^{I}$ are the leading twist
quark $2\pi$DAs with isospin $I$.
For general properties of the two-pion distribution amplitudes see
the next section.

The leading twist skewed quark distribution
with $f=f'=q=u,d,s$ and $B_1=B_2=$ proton (Eq.~(\ref{SPD1}))
can be decomposed into nucleon
spin non-flip and spin flip parts.
Here we adopted the notations of Ji \cite{JiReview} for the spin decomposition
of the matrix element of a bilocal quark operator between proton states.
The spin non-flip part is denoted by $H_f$,
the spin flip part by $ E_f$. They are defined by:
\be
\nonumber
\int \frac{d\lambda }{2\pi }e^{i\lambda \tau}\langle p^{\prime }|\bar
\psi_f
(-\lambda n/2){\hat n}
 \psi_f (\lambda n/2)|p\rangle
&=&  H_f(\tau,\xi ,t) \;
\bar U(p^{\prime }) \; \hat n  \; U(p) \\
&+& \frac 1{2M_N} \;  E_f(\tau,\xi ,t) \;
\bar U(p^{\prime }) \;\sigma_{\mu\nu} n^\mu  \Delta^\nu  U(p) \, .
\label{E-H-QCD-2}
\ee
The independent variables of the skewed quark distributions
$ H_q(\tau,\xi ,\Delta ^2)$ and $ E_q(\tau,\xi ,\Delta ^2)$ are chosen
to be
$\tau$, related to the fraction of the target momentum
carried by the interacting parton $x_1$ by:
\be
\nonumber
\tau=\frac{2\ x_1-x}{2-x}\, ,
\ee
 the square of the four-momentum transfer $\Delta^2=t$, and
the skewedness parameter $\xi$ defined in Eq.~(\ref{xi-def})%
\footnote{The skewed parton distributions
and meson distribution amplitudes are scale dependent.
The scale is set by the photon
virtuality $Q^2$. We do not show the scale dependence of
these quantities to simplify the notation.}.
In the forward case $p = p'$ both $\Delta$ and $\xi$ are zero and
the second term on the r.h.s.\ of Eq.~(\ref{E-H-QCD-2}) disappears. The
function $ H$ becomes the usual parton distribution
function:
\be
 H_q(\tau, \xi = 0, t= 0) &=& q(\tau)
=\left\{
\begin{array}{cr}
q(\tau),& \hspace{.5cm} \tau \; > \; 0\,, \\
-\bar q(-\tau)
,& \hspace{.5cm} \tau \; < \; 0 \, .
\end{array}
\right.
\label{forward_limit}
\ee
Using the symmetry properties for the $2\pi$DAs (\ref{reflection}) the leading
order amplitude for $\pi^+\pi^-$ production can be written as:
\be
\nonumber
&&\langle p(p'), \pi^+\pi^-(q')| J^{\rm e.m.}\cdot
\varepsilon_L|p(p)\rangle=
-(e 4 \pi \alpha_s)\frac{N_c^2-1}{N_c^2} \frac{1}{24 Q} \\
\nonumber
&\times&
\Biggl[
\int_0^1 dz
\frac{(1-2 z)\Phi^{I=0}(z,\zeta,m_{\pi\pi})
-\frac{2}{C_F} \Phi^{G}(z,\zeta,m_{\pi\pi})}{z(1-z)}\\
\nonumber
&\times&\Biggl\{ \bar U(p')\hat n  U(p)
\int_{-1}^1 d\tau
\Bigl(2\ H_u(\tau,\xi,t)- H_d(\tau,\xi,t)\Bigl) \alpha^-(\tau) \\
\nonumber
&+&\bar U(p')
\frac{\sigma_{\mu\nu} n^\mu\Delta^\nu}{2 M_N}  U(p)
\int_{-1}^1 d\tau
\Bigl(2\ E_u(\tau,\xi,t)- E_d(\tau,\xi,t)\Bigl) \alpha^-(\tau)
\Biggl\}  \\
\nonumber
&+&\int_0^1 dz \frac{\Phi^{I=1}(z,\zeta,m_{\pi\pi})}{z(1-z)}
\Biggl\{ \bar U(p')\hat n  U(p)
\int_{-1}^1 d\tau
\Bigl(2\ H_u(\tau,\xi,t)+ H_d(\tau,\xi,t)\Bigl) \alpha^+(\tau) \\
&+&\bar U(p')
\frac{\sigma_{\mu\nu} n^\mu\Delta^\nu}{2 M_N}  U(p)
\int_{-1}^1 d\tau
\Bigl(2\ E_u(\tau,\xi,t)+  E_d(\tau,\xi,t)\Bigl) \alpha^+(\tau)
\Biggl\} \Biggl]   \, ,
\label{ampp}
\ee
where
\be
\alpha^{\pm}(\tau)=\frac{1}{\tau+\xi-i0}\pm
                \frac{1}{\tau-\xi+i0}\, .
\ee
Let us note that for the production of two pions in an isovector state
there is an additional contribution proportional to the skewed gluon
distribution in the nucleon, see Fig.~\ref{graph}c).
The corresponding expression
is given by the last term of Eq.~(\ref{P-general}) and also
can be found in
\cite{Radyushkinrho,MPW}. Such a contribution is absent if the pions
are produced in
the isoscalar state. In the present paper
we are interested in two pion production in the valence region and in
particular in the isospin zero state in order to identify the gluons of
Fig.~\ref{graph}b). There
the contribution from gluons in the nucleon can be safely neglected
\cite{Vander}.

For the $\pi^0\pi^0$ production amplitude we have:
\be
\nonumber
&&\langle p(p'), \pi^0\pi^0(q')| J^{\rm e.m.}\cdot
\varepsilon_L|p(p)\rangle=
-(e 4 \pi \alpha_s)\frac{N_c^2-1}{N_c^2} \frac{1}{24 Q} \\
\nonumber
&\times&
\int_0^1 dz \frac{(1-2 z)\Phi^{I=0}(z,\zeta,m_{\pi\pi})
-\frac{2}{C_F} \Phi^{G}(z,\zeta,m_{\pi\pi})}{z(1-z)}\\
\nonumber
&\times&\Biggl\{ \bar U(p')\hat n  U(p)
\int_{-1}^1 d\tau
\Bigl(2\ H_u(\tau,\xi,t)- H_d(\tau,\xi,t)\Bigl) \alpha^-(\tau) \\
&+&\bar U(p')
\frac{\sigma_{\mu\nu} n^\mu\Delta^\nu}{2 M_N}  U(p)
\int_{-1}^1 d\tau
\Bigl(2\ E_u(\tau,\xi,t)- E_d(\tau,\xi,t)\Bigl) \alpha^-(\tau)
\Biggl\} \, .
\label{amp0}
\ee
The expression for the ratio of the
$I=0$ and $I=1$ amplitude was obtained recently by Diehl et~al.~\cite{DGP99},
however, the authors missed the contribution of the gluon $2\pi$DA
(Fig.~\ref{graph}b)).
Below we shall see that the contribution of this amplitude to the total
two-pion amplitude is larger than the contribution of the quark $2\pi$DAs.

\vspace{0.3cm}
\noindent{\bf Two pion distribution amplitudes}
\vspace{0.2cm}\\
The two-pion distribution amplitudes describe the fragmentation of a pair of
collinear partons (quarks or gluons) into the final pion pair \cite{Ter}.
As we shall discuss in the following, they can be related
by crossing to the ordinary parton distributions in a pion.
From C-parity one can easily derive the following symmetry
properties of the $2\pi$DAs:
\be
\nonumber
\Phi^{I=0}(z,\zeta,m_{\pi\pi})&=&
-\Phi^{I=0}(1-z,\zeta,m_{\pi\pi})=\Phi^{I=0}(z,1-\zeta,m_{\pi\pi}),\\
\nonumber
\Phi^{I=1}(z,\zeta,m_{\pi\pi})&=&\Phi^{I=1}(1-z,\zeta,m_{\pi\pi})=
-\Phi^{I=1}(z,1-\zeta,m_{\pi\pi})\, ,\\
\Phi^{G}(z,\zeta,m_{\pi\pi})&=&
\Phi^{G}(1-z,\zeta,m_{\pi\pi})=\Phi^{G}(z,1-\zeta,m_{\pi\pi})\, .
\label{reflection}
\ee

Following \cite{KMP99,MVP98}
we decompose both quark and gluon two-pion distribution amplitudes in
conformal and partial waves. For quark $2\pi$DAs the decomposition reads: \be
\nonumber
\Phi^{I=0}(z, \zeta, \mpp^2 )&=&6z(1-z)
\sum_{\scriptstyle n=1 \atop \scriptstyle {\rm odd}}^{\infty}
\sum_{\scriptstyle l=0 \atop \scriptstyle {\rm even}}^{n+1}
B_{nl}^{I=0}(\mpp) C_n^{3/2}(2 z-1)
P_l(2\zeta-1),\\
\Phi^{I=1}(z, \zeta, \mpp^2 )&=&6z(1-z)
\sum_{\scriptstyle n=0 \atop \scriptstyle {\rm even}}^{\infty}
\sum_{\scriptstyle l=1 \atop \scriptstyle {\rm odd}}^{n+1}
B_{nl}^{I=1}(\mpp) C_n^{3/2}(2 z-1) P_l(2\zeta-1),
\label{razhlq}
\ee
while for gluon $2\pi$DAs we have:
\be
\Phi^G(z, \zeta, \mpp^2 )=
30\ z^2(1-z)^2
\sum_{\scriptstyle n=0 \atop \scriptstyle {\rm even}}^\infty
\sum_{\scriptstyle l=0 \atop \scriptstyle {\rm even}}^{n+2}
A^G_{nl}(\mpp) C^{5/2}_n(2 z-1) P_l(2\zeta-1)\, .
\label{razlg}
\ee
Here $P_l(x)$ are the Legendre polynomials and $C_n^\mu(x)$
are Gegenbauer polynomials.

The distribution amplitudes
are constrained by soft pion theorems and crossing
relations.  The soft pion theorems for $2\pi$DAs read \cite{KMP99,MVP98}:
\be
\nonumber
\Phi^{I=0} (z, \zeta= 1, \mpp= 0 )&=&
\Phi^{I=0} (z, \zeta= 0, \mpp= 0 )=0\, ,\\
\nonumber
\Phi^{I=1} (z, \zeta= 1, \mpp= 0 )&=&
-\Phi^{I=1} (z, \zeta= 0, \mpp= 0 )=\varphi(z)\, ,\\
\Phi^G (z, \zeta= 1, \mpp= 0 )&=&
\Phi^G (z, \zeta= 0, \mpp= 0 )=0\, .
\label{sptg}
\ee
Here $\varphi(z)$ is the pion distribution amplitude.

Additional constraints for the quark and gluon $2\pi$DAs are provided by the
crossing relations between two-pion distribution amplitudes and the known
parton distributions in the pion. For the derivation of such relations for
quark $2\pi$DAs see \cite{MVP98}, for the gluon $2\pi$DA see
\cite{KMP99}. The crossing relations have the form:
\be
\ba{l}
\displaystyle
B_{N-1,N}^{I=0}(0)=
\frac 23\ \frac{2N+1}{N+1} \int_{0}^1 dx\ x^{N-1}
\frac {1}{N_f}\sum_q(q_\pi(x)+\bar q_\pi(x))\, ,
\\[8mm] \displaystyle
B_{N-1,N}^{I=1}(0)=
\frac 23\ \frac{2N+1}{N+1} \int_{0}^1 dx\ x^{N-1}
(u_\pi(x)-\bar u_\pi(x))\, ,
\\[8mm] \displaystyle
A_{N-2,N}^G(0)=\frac{4}{5}\
\frac{2N+1}{(N+1)(N+2)}
\int_{0}^1 dx\ x^{N-1}
g_\pi(x)\, ,
\ea
\label{crossing}
\ee
where $q_\pi(x)$, $\bar q_\pi(x)$, and $g_\pi(x)$ are the usual quark,
antiquark, and gluon distributions in the pion.
Using these relations one
can easily derive the normalization of the two-pion distribution
amplitude ($2\pi$DA) at $\mpp=0$:
\be
\ba{l}
\displaystyle
\int_0^1 dz (2 z-1)\Phi^{I=0}(z, \zeta, \mpp=0 )
=-\frac{4}{N_f}\ M_2^{Q}\zeta(1-\zeta) \, ,
\\[4mm] \displaystyle
\displaystyle
\int_0^1 dz \Phi^{I=1}(z, \zeta, \mpp )
=(2\zeta-1) F_\pi(\mpp) \, ,
\\[4mm] \displaystyle
\int_0^1 dz\ \Phi^G (z, \zeta, \mpp=0)=
-2\ M_2^G\zeta(1-\zeta) \, .
\ea
\ee
Here $M_2^Q$ and $M_2^G$ are momentum fractions carried
by quarks or gluons respectively in the pion and $F_\pi(\mpp)$
is the pion e.m.~form factor.
The soft pion theorems and crossing relations constrain
$\Phi^{I=0}(z,\zeta,\mpp)$ and $\Phi^G (z, \zeta, \mpp )$
only at $\mpp=0$. For higher $\mpp$ one can apply the dispersion
relation analysis of Ref.~\cite{MVP98}.

The asymptotic expression for the isovector $2\pi$DA has the form \cite{PW98}:
\be
\Phi^{I=1}_{asy} (z, \zeta, \mpp )=
6\ z(1-z)\ (2 \zeta-1) F_\pi(\mpp)\, .
\label{asy1}
\ee
Under evolution $\Phi^G$
and $\Phi^{I=0}$ mix with each other \cite{mix}.
For asymptotically large $Q^2$ one obtains \cite{KMP99}:
\be
\ba{l}
\displaystyle
\Phi^G_{asy} (z, \zeta, \mpp=0 )=
- \frac{240\ C_F}{N_f+4 C_F}\ z^2(1-z)^2 \ \zeta(1-\zeta)\, ,
\\[4mm]  \displaystyle
\Phi^{I=0}_{asy} (z, \zeta, \mpp =0 )=
-\frac{120}{N_f+4 C_F}\ z(1-z)\ (2 z-1) \ \zeta(1-\zeta)\, .
\ea
\label{asymptotic}
\ee
In the following we will use the asymptotic expressions for
the $2\pi$DAs.

\vspace{0.3cm}
\noindent{\bf Numerical results and discussion}
\vspace{0.2cm}\\
Using our results for the production amplitudes discussed in the
previous sections we can easily estimate the ratio of the cross sections
for hard exclusive two pions production in the isoscalar and isovector
states. The amplitudes of
physically accessible reactions can be expressed in terms of
$I=0$ and $I=1$ amplitudes as follows:
\be
\nonumber
M^{\pi^+\pi^-}&=&M^{I=0}+M^{I=1}\, ,\\
M^{\pi^0\pi^0}&=&M^{I=0}\, .
\ee
For the ratio
\be
R(x_{Bj},\mpp,t)=\frac{|M^{I=0}|^2}{|M^{I=I}|^2}
\label{ratio}
\ee
(integrated over the scattering angle of the pions) the $x_{Bj}$, $t$,
and $\mpp$ dependence factorizes in the leading order:
\be
R(x_{Bj},\mpp,t)&=&R_{\rm DA}(\mpp) R_{\rm SPD}(x_{Bj},t)\, ,
\ee
where the $\mpp$ dependence is governed by the $2\pi$DAs:
\be
R_{\rm DA}(\mpp)&=&\frac{D^{I=0}(\mpp)}{D^{I=1}(\mpp)}\,
\ee
with
\be
\nonumber
D^{I=0}(\mpp)&=&\beta
\int_{-1}^1 d \cos \theta
\Biggl|
\int_0^1 dz
\frac{(1-2 z)\Phi^{I=0}(z,\zeta,m_{\pi\pi})
-\frac{2}{C_F} \Phi^{G}(z,\zeta,m_{\pi\pi})}{z(1-z)}
\Biggr|^2\, ,\\
D^{I=1}(\mpp)&=&\beta
\int_{-1}^1 d \cos \theta
\Biggl|
\int_0^1 dz
\frac{\Phi^{I=1}(z,\zeta,m_{\pi\pi})}{z(1-z)}\Biggr|^2\,.
\label{RDA}
\ee
Here $\theta$ is the scattering angle of the pions in their c.m.
frame relative to the direction of $q'$:
\be
\beta \cos \theta =2\zeta-1 \ \ {\rm with\ }\
\beta=\sqrt{1-\frac{4 m_\pi^2}{\mpp^2}}\, .
\ee
The dependence on Bjorken $x=2\xi/(1+\xi)$ and the squared momentum transfer
$t$ is determined by integrals over the skewed parton distributions (SPD):
\be
R_{\rm SPD}(x_{Bj},t)=\frac{|2I^-_u(\xi,t)-I^-_d(\xi,t)|^2}{|2I^+_u(\xi,t)+
I^+_d(\xi,t)|^2}
\ee
with
\be
I_f^\pm(\xi,t)=\int_{-1}^1 d\tau H_f(\tau,\xi,t) \Biggl[
\frac{1}{\tau+\xi-i 0} \pm \frac{1}{\tau-\xi+i 0}
\Biggr]\, .
\ee

\begin{figure}[t]
\vspace*{1cm}
\centering
\epsfig{file=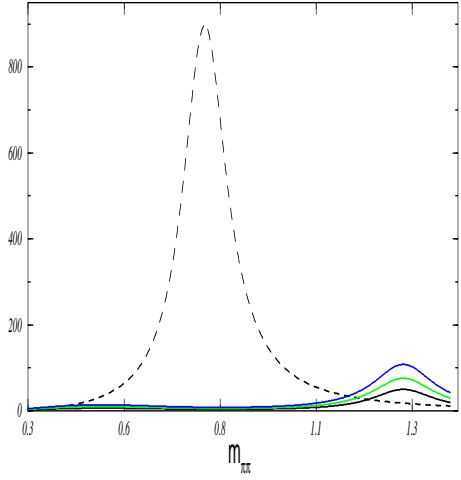, height=6.5cm, width=6cm}
\caption{{\it The shape of
two-pion mass $\mpp$ (\mbox{GeV})
distributions
for pions with isospin one (dashed curve) and isospin zero
(solid curves). The isospin zero distributions are plotted for
Bjorken $x=0.3,0.4,0.5$. (The larger $x_{Bj}$ the more enhanced is
the distribution.) } }
\label{fig1}
\end{figure}

For the quark skewed parton distributions in the integrals
$I_f^\pm(\xi,t)$ we used the models resulting from Radyushkin's double
distributions \cite{Rad}. They are obtained assuming
$F(x,y)=\pi(x,y)\cdot q(x)$ with the profile function
$\pi(x,y)=\frac{6y(1-x-y)}{(1-x)^3}$ suggested in \cite{Rad2} and the MRS(A')
parameterizations for the usual (forward) parton distributions $q(x)$
\cite{MRS}. The SPDs $H_f(\tau,\xi)$ are related to Radyushkin's
nonforward parton distributions ${\cal F}_{\zeta}^f(X)$ by:
\be
H_f(\tau,\xi)=\frac{1}{1+\xi}\left(\Theta(\tau+\xi)
{\cal F}^{f}_{\zeta}\Big(\frac{\tau+\xi}{1+\xi}\Big)
-\Theta(\xi-\tau)
{\cal F}^{\bar f}_{\zeta}\Big(\frac{\xi-\tau}{1+\xi}\Big)\right)
\ee
with $\zeta=\frac{2\xi}{1+\xi}$, while the nonforward parton distributions
are obtained from the double distributions by the integral
\be
{\cal F}_{\zeta}(X)
=\int_{0}^{\mbox{min}\{\frac{X}{\zeta},\frac{(1-X)}{(1-\zeta)}\}}
F(X-\zeta y,y)dy\, .
\ee
For the $t$-dependence we use the factorized ansatz
$H(\tau,\xi,t)=H(\tau,\xi)F_1(t)$ with the proton form factor $F_1(t)$.
For simplicity we do not consider the specific contributions to
quark SPDs which are not taken into account by double distributions.
These contributions
were observed first in chiral quark soliton model calculations
\cite{ofpd} but actually they are of quite general nature as was shown
in \cite{PW99}. Also we neglected the contribution of the nucleon
spin flip skewed parton distributions $E$. The complete
studies, involving all these refinements, will be published elsewhere.

Using the asymptotic expressions for the $2\pi$DAs
(\ref{asy1},\ref{asymptotic})
we can compute the integrals over $z$ entering Eq.~(\ref{RDA}):
\be
\nonumber
\int_0^1 dz
\frac{(1-2 z)\Phi^{I=0}(z,\zeta,m_{\pi\pi})
}{z(1-z)}
&=&\frac{40}{N_f+4C_F}\Bigl\{
\frac{3C-\beta^2}{12}\ f_0(\mpp)\ P_0(\cos\theta) \\
\label{WFint}
&&-
\frac{\beta^2}{6} \ f_2(\mpp) \ P_2(\cos\theta)
\Bigr\}\, ,  \\
\nonumber
-\frac{2}{C_F}
\int_0^1 dz
\frac{
 \Phi^{G}(z,\zeta,m_{\pi\pi})}{z(1-z)}
&=&\frac{80}{N_f+4C_F}\Bigl\{
\frac{3C-\beta^2}{12}\ f_0(\mpp)\ P_0(\cos\theta) \\
&&-
\frac{\beta^2}{6} \ f_2(\mpp) \ P_2(\cos\theta)
\Bigr\}\, , \\
\int_0^1 dz
\frac{\Phi^{I=1}(z,\zeta,m_{\pi\pi})
}{z(1-z)}
&=&6\ \beta \ F_\pi(\mpp)\ P_1(\cos\theta) \, .
\ee
(The effects due to deviations from the asymptotic forms were considered
in \cite{MVP98,DIS99}). We see that for the production
of the pion pair in the isoscalar state the contribution of the gluon
$2\pi$DA is two times larger than the contribution of the quark $2\pi$DAs.
Therefore the hard production of the pions in the isoscalar
state mainly probes the gluon structure of the pion.

For our estimates we use the parameterization of
the pion electromagnetic form factor $F_\pi(\mpp)$ given
in Ref.~\cite{guerro}. This parameterization reproduces the
results of two-loop chiral perturbation theory at small $\mpp$
and is in very good agreement with experimental data \cite{barkov}.

The functions $f_{0,2}(W)$, the so-called Omn\`es functions \cite{omnes},
can be related to the $\pi\pi$ phase shifts $\delta_0^0(W)$ and
$\delta_2^0(W)$ using Watson's theorem \cite{Watson} and the dispersion
relations derived in \cite{MVP98}:
\be
f_l(\mpp)=
\exp\biggl[i\delta_l^0(\mpp)+
\frac{\mpp^{2}}{\pi}
{\rm Re} \int_{4m_\pi^2}^\infty
ds \frac{\delta_l^0(s)}{s(s-\mpp^2-i0)}
\biggr]\, .
\label{omnes}
\ee
The Omn\`es functions were analyzed in detail in Ref.~\cite{Gasser}
and for the estimates of the cross sections one can use the results
of that work.

The constant $C$ in Eqs.~(\ref{WFint}) plays the role of an integration
constant in the Omn\`es solution to the corresponding dispersion relation.
From the soft pion theorem it follows that $C=1+O(m_\pi^2)$.
Using the instanton model for calculations
of $B_{nl}(\mpp)$ at low energies
\cite{MVP98,PW98} one finds \cite{KMP99}
the constant
$C$ to be equal to:
\be
\nonumber
C=1+b\ m_\pi^2 + O(m_\pi^4)\ {\rm \ with \ \ \ \ }
b\approx -1.7\ {\rm GeV}^{-2}\, .
\ee

The Omn\`es functions contain information about the  $\pi\pi$
resonances as well as about the nonresonant background. Due to the
$f_2(1270)$ resonance the $f_2(\mpp)$ has a peak at $\mpp=1.275$~GeV.
For the numerical calculation we use the Pad\'e approximation for
the Omn\`es function $f_0(\mpp)$ suggested in Ref.~\cite{f0}.
For estimates of $f_2(\mpp)$ we used the expression (\ref{omnes})
where the D-wave pion scattering phase shift we saturated by
the contribution of the $f_2(1270)$ resonance.

\begin{figure}[t]
\vspace*{1cm}
\centering
\epsfig{file=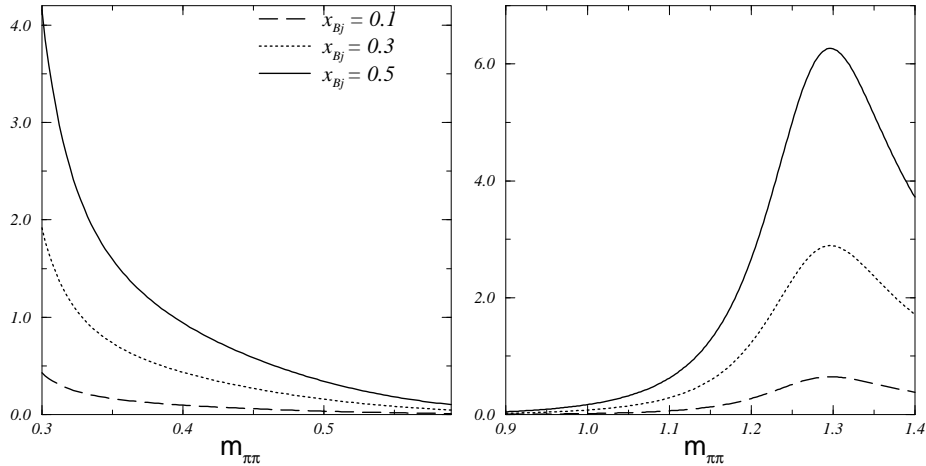, height=6.5cm}
\caption{{\it The ratio $\frac{dN^{I=0}}{d\mpp}/\frac{dN^{I=1}}{d\mpp}$
in two regions of $\mpp$, around $0.4$ \mbox{GeV} (left panel)
and around $1.3$ \mbox{GeV} (right panel).}}
\label{fig2}
\end{figure}

In Fig.~\ref{fig1} we plot the two-pion mass distributions in the isoscalar
and isovector channel at different values of Bjorken $x$.

The two-pion mass distribution in the isovector channel at large $Q^2$
is determined by the pion e.m.~form factor \cite{MVP98}:
\be
\frac{dN^{I=1}}{d\mpp}\propto 36\ \beta^3\ \mpp\ |F_\pi(\mpp)|^2\,,
\ee
which is plotted in Fig.~\ref{fig1} as a dashed curve.
For the two-pion mass distribution in the isoscalar channel (for
$t=t_{\rm min}$) we can use:
\be
\frac{dN^{I=0}}{d\mpp}=R(x_{Bj},\mpp,t_{\rm min})\ \frac{dN^{I=1}}{d\mpp}\, ,
\ee
where $R(x_{Bj},\mpp,t)$ is given by Eq.~(\ref{ratio}).
In Fig.~\ref{fig1} we plot $\frac{dN^{I=0}}{d\mpp}$
(solid lines) for $x_{Bj}=0.3,0.4,0.5$. We see that this
distribution is enhanced around $\mpp=0.4$~GeV due to the
S-wave contribution and around $\mpp=1.3$~GeV due to D-waves.
To illustrate the enhancements in these regions we plot
in Fig.~\ref{fig2} the ratio $\frac{dN^{I=0}}{d\mpp}/\frac{dN^{I=1}}{d\mpp}$
at different values of Bjorken $x$.
From these figures we see that the most favorable regions to
observe isoscalar channel in $\pi^+\pi^-$ productions
(and hence the gluonic component of $2\pi$DA)
are $x_{Bj}>0.3$ and $\mpp$ around 0.4~GeV (S-wave) and 1.3~GeV
(D-wave).

\begin{figure}[t]
\vspace*{1cm}
\centering
\epsfig{file=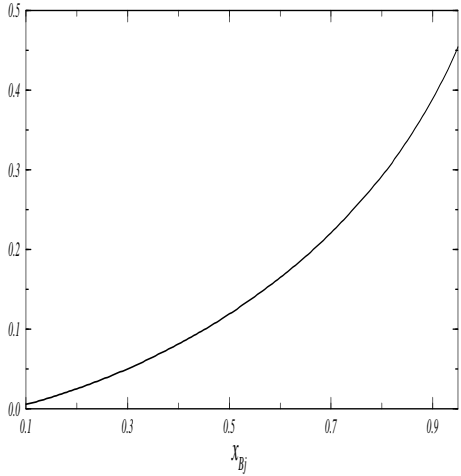, height=6.5cm, width=6cm}
\caption{{\it The ratio of
$\frac{d \sigma^{\pi^0\pi^0}}{dt}|_{t=t_{\rm min}}$
to
$\frac{d \sigma^{\pi^+\pi^-}}{dt}|_{t=t_{\rm min}}$
integrated over $\mpp$ from the threshold to 1.4 GeV as
 a function of Bjorken $x$.} }
\label{fig3}
\end{figure}
On Fig.~\ref{fig3} we show the estimates of the ratio of
$\frac{d \sigma^{\pi^0\pi^0}}{dt}|_{t=t_{\rm min}}$
to
$\frac{d \sigma^{\pi^+\pi^-}}{dt}|_{t=t_{\rm min}}$ integrated
over $\mpp$ from the threshold $\mpp=2 m_\pi$ to $\mpp=1.4$~GeV.
We see that in the valence region $x_{Bj}=0.1$ -- $0.4$ the
cross section for $\pi^0\pi^0$ production is a few percent compared to
that for $\pi^+\pi^-$  production
(which is usually called $\rho$ production).

Let us note that the most promising way to determine the
$I=0$ production amplitude is through its interference
with the $I=1$ amplitude. Experimentally this can be done
studying the angular distributions of the produced
$\pi^+\pi^-$ pairs.  For example, the intensity density
$\langle \cos \theta \rangle$ ($\theta$
is the polar angle of the pion momentum in the CM frame with respect to
the direction of the pion pair momentum $q'$)
directly probes the interference
of the $I=0$ and the $I=1$ channel. It would be zero if the $\pi^+\pi^-$
pairs were produced purely in the $I=1$ state. So the observation
of a nonzero $\langle \cos \theta \rangle$ would clearly
indicate the production of pion pairs in the isoscalar state.
The corresponding expressions
can be obtained easily using Eqs.~(\ref{ampp},\ref{WFint}).
The full expression is rather tedious, here we give
the simplified expression
for the intensity density $\langle \cos \theta \rangle$ at $\mpp$
close to the mass of the $\rho$ meson:
\be
\langle \cos \theta \rangle \approx
\frac{4}{3 |F_\pi(m_\rho)|}\ \frac{1}{N_f+4\ C_F}
{\rm Im}\Biggl[
\frac{2I_u^- - I_d^-}{2I_u^+ + I_d^+}\
\biggl(5 f_0(m_\rho)-
2 f_2(m_\rho)\biggr)
\Biggr]\, .
\ee
Numerically we get for $\langle \cos \theta \rangle$
values of about -4\%, -7\%, and -10\%
at $\mpp=m_\rho$ and $x_{Bj}=0.1, 0.2,$ and 0.3 respectively.
At lower invariant masses $\langle \cos \theta \rangle$
is larger. At $\mpp=0.5$~GeV for example the values of
$\langle \cos \theta \rangle$ are -18\%, -27\% and -33\%
for $x_{Bj}=0.1, 0.2,$ and 0.3 respectively.
Detailed estimates of various intensity
densities will be done elsewhere.

\vspace{0.3cm}
\noindent{\bf Summary}\nopagebreak
\vspace{0.2cm}\\
We have calculated hard exclusive two-pion production at the leading twist
level to leading order in $\alpha_s$. The amplitude is expressed in terms of
skewed parton distributions (SPD) and two-pion distribution amplitudes
($2\pi$DA).  For the $2\pi$DAs
we used the asymptotic expressions.
The comparison of the cross sections
for the production of isospin one and isospin zero pion
pairs shows that for small Bjorken $x$ ($x_{Bj}<0.1$) the isospin-one channel
dominates, whereas in the valence region ($x_{Bj}>0.3$) at certain invariant
two-pion masses the contribution of isospin zero production is also sizeable.
We showed that the isospin zero amplitude is dominated by the gluon $2\pi$DA.
Hence the cross section for isospin zero pion pairs
gives access to the gluon distribution in the pions. To extract the isospin
zero channel measurements in the quark valence region are needed.
This extraction is facilitated if $\pi^0\pi^0$ pairs can be detected
since the isospin one channel contributes only to $\pi^+\pi^-$ pairs.

Let us note that the general expression for the leading twist
amplitude Eq.~(\ref{P-general}) can be used to study other channels
like $K \bar K$, $\pi K$, etc. By crossing symmetries this allows to
obtain information on the partonic distributions of the produced particles.
This is particularly important when the particles are not available as targets
in deep inelastic reactions. The exclusive two-particle production can be
accessible for measurements e.g. in the COMPASS experiment \cite{compass}.

We are aware of the fact that higher twist corrections could be
large. They have not been analyzed yet but there is some hope
that they cancel at least partially in the ratios of cross sections
we discussed.

\vspace{0.3cm}
\noindent{\bf Acknowledgments}
\vspace{0.2cm}\\
We are grateful to B.~Clerbaux,
N.~Kivel, L.~Mankiewicz, M.~Penttinen,
E.~Stein, M.~Vanderhaeghen,
and C.~Weiss for useful discussions. Special thanks are due to
L.~Frankfurt and M.~Strikman for encouragement to look into the
problem and for numerous discussions. The work has been partially supported by
the Russian-German exchange program, Studienstiftung des deutschen Volkes,
DFG, BMBF, and COSY-Juelich.

\end{document}